\documentclass[conference]{IEEEtran}
\IEEEoverridecommandlockouts

\usepackage{amsmath,amssymb,amsfonts}
\usepackage{graphicx}
\usepackage{textcomp}

\usepackage{booktabs}
\usepackage{tabularx}
\usepackage{float}
\usepackage{multirow}
\usepackage{xcolor}
\usepackage{tikz}
\usepackage{url}

\usepackage[
 backend=biber
,style=ieee
,minbibnames=1
,maxbibnames=2
,mincitenames=1
,maxcitenames=2
,eprint=true
,doi=true
,isbn=false
,url=true
]{biblatex}
\bibliography{../ref.bib} 
\AtBeginBibliography{\footnotesize}

\DeclareSourcemap{
  \maps{
    \map{
      \step[fieldset=month, null]
      \step[fieldset=address, null]
      \step[fieldset=location, null]
      \step[fieldset=publisher, null]
      \step[fieldset=url, null]
      \step[fieldset=isbn, null]
      \step[fieldset=editor, null]
    }
  }
}

\usepackage[
hidelinks,
  colorlinks=true,
  citecolor=blue,
  pdfstartview=Fit,
  breaklinks=true
]{hyperref}

\usepackage[capitalise,noabbrev]{cleveref}
\crefname{figure}{Fig.}{Figs.}
\Crefname{figure}{Fig.}{Figs.}

\newcommand{\RFFlowFeaturesAccuracy}{0.9680}
\newcommand{\RFFlowFeaturesBalancedAccuracy}{0.8427}

\newcommand{\XGBFlowFeaturesBalancedAccuracy}{0.9271}

\newcommand{\RFSPLTBalancedAccuracy}{0.6040}

\newcommand{\XGBSPLTBalancedAccuracy}{0.7534}

\newcommand{\CNNOneDSPLTAccuracy}{0.9682}
\newcommand{\CNNOneDSPLTBalancedAccuracy}{0.9753}
\newcommand{\CNNOneDSPLTMacroFOne}{0.8932}
\newcommand{\CNNOneDSPLTWeightedFOne}{0.9705}

\begin{document}

\title{Matched-View Cross-Domain Evaluation of WireGuard VPN Traffic Classification Using Early-Flow Fingerprints}

\author{
  \IEEEauthorblockN{Yasameen Sajid Razooqi\IEEEauthorrefmark{1} and
                    Adrian Pekar\IEEEauthorrefmark{1}\IEEEauthorrefmark{2}}
  \IEEEauthorblockA{\IEEEauthorrefmark{1}Budapest University of Technology and Economics, Budapest, Hungary}
  \IEEEauthorblockA{\IEEEauthorrefmark{2}CUJO LLC Hungary}
  \IEEEauthorblockA{E-mail: \{rsajid, apekar\}@hit.bme.hu}
}

\maketitle

\begin{tikzpicture}[remember picture,overlay]
\node[anchor=north, align=center, font=\scriptsize, text width=.92\paperwidth,
      yshift=-.3cm] at (current page.north) {%
  \copyright~2026 IEEE. Personal use of this material is permitted. Permission
  from IEEE must be obtained for all other uses, in any current or future media,
  including reprinting/republishing this material for advertising or promotional
  purposes, creating new collective works, for resale or redistribution to
  servers or lists, or reuse of any copyrighted component of this work in other
  works.\\
  Accepted for presentation at the 22nd International Conference on Network and
  Service Management (CNSM 2026).};
\end{tikzpicture}

\begin{abstract}
Classifying VPN-encrypted traffic by application category typically relies on datasets that collect non-VPN and VPN traffic in separate sessions, conflating encapsulation effects with session-level differences in user behavior, timing, and application mix. We use a recently published WireGuard tunnel dataset in which pre- and post-tunnel traffic is captured simultaneously, with a packet-level match ratio above 99.9\%. This matched-capture design eliminates session-level confounds and enables a cross-domain benchmark: models are trained on non-VPN flows and tested on the VPN view of the same underlying flows. We compare whole-flow statistical aggregates (FlowFeatures) and Sequence of Packet Length and Time (SPLT) early-flow fingerprints across Random Forest, XGBoost, and a multi-scale CNN1D. Cross-domain transfer depends jointly on representation and model: tree ensembles achieve balanced accuracy of 0.84--0.93 with FlowFeatures but only 0.60--0.75 with flattened SPLT, whereas CNN1D processes the same SPLT fingerprint as a sequence and achieves the strongest transfer overall (balanced accuracy 0.98, macro F1 0.89) without any VPN data during training.
\end{abstract}


\section{Introduction}
\label{sec:introduction}

Classifying encrypted traffic by application category is essential for network management, quality-of-service enforcement, and security monitoring. VPN tunnels complicate this task: encryption removes payload-level cues, and encapsulation distorts the packet sizes and inter-arrival times that flow-level classifiers rely on~\cite{11091298}. At the deployment level, VPN tunnels may multiplex traffic from multiple applications, though this paper focuses on the per-flow classification problem.

A key challenge in evaluating VPN traffic classifiers is the lack of datasets that provide both unencapsulated and VPN-encapsulated views of the same underlying flows. Existing public benchmarks collect non-VPN and VPN traffic in separate sessions, meaning that any performance degradation observed in cross-domain transfer may reflect session-level differences in user behavior, timing, and application mix rather than the effect of encapsulation alone. Without controlling for these confounds, it is difficult to determine how much classification-relevant signal actually survives VPN encapsulation and which feature representations and model architectures are best suited to exploit it.

This paper addresses that gap using a recently published WireGuard tunnel dataset~\cite{Razooqi2026Pekar} in which pre-tunnel and post-tunnel traffic is captured simultaneously, with inner packets matched to their encrypted outer counterparts at a ratio above 99.9\%. Application labels are assigned via deep packet inspection on the inner side and transferred to the outer view through this matching. Because both views describe the same underlying flows at the same time, the session-level confound introduced by separately collected runs does not exist, enabling a direct measurement of what encapsulation does to the classification-relevant signal. The dataset construction is documented in the cited data article; the contribution here is the matched-view evaluation protocol and the resulting representation--model evidence.

We evaluate two feature representations, whole-flow statistical aggregates (FlowFeatures) and Sequence of Packet Length and Time (SPLT) early-flow fingerprints, paired with three classification models: Random Forest, XGBoost, and a multi-scale 1D convolutional network (CNN1D). Models are trained exclusively on non-VPN flows and tested on the matched VPN view of the same underlying flows. The central finding is that classical models achieve substantially lower balanced accuracy with flattened SPLT than with FlowFeatures, whereas CNN1D processes SPLT without discarding its sequential structure and leads every aggregate metric (balanced accuracy 0.98, macro F1 0.89) without any exposure to VPN data during training. Our results reveal that cross-domain transfer depends jointly on the feature representation and the model family.

The contributions of this paper are:
\begin{itemize}
  \item A matched-view cross-domain transfer benchmark based on simultaneous pre- and post-VPN captures of the same underlying flows, which eliminates the session-level confounds present in prior datasets that collect VPN and non-VPN traffic separately.
  \item Empirical evidence that SPLT early-flow fingerprints transfer substantially better under a multi-scale CNN than under tree-based classifiers that flatten the sequence.
  \item Evidence that cross-domain transfer success depends jointly on the feature representation and the model family: tree ensembles achieve higher balanced accuracy and macro F1 with FlowFeatures, while CNN1D achieves the strongest result with SPLT.
\end{itemize}

\section{Related Work}
\label{sec:related}

\paragraph{Cross-domain and VPN-aware classification}
Several recent methods address encrypted traffic classification under VPN conditions, with varying degrees of cross-domain evaluation. \citeauthor{Shapira2021}~\cite{Shapira2021} convert flows into two-dimensional packet-size and timing histograms (FlowPic) and show that this visual representation can preserve category structure when traffic is observed through VPN or Tor tunnels. \citeauthor{MengJuKuo2023}~\cite{MengJuKuo2023} arrange bidirectional flow statistics into a semantically ordered token matrix and evaluate CNN, BiLSTM, and Transformer models on ISCX-VPN-2016, reporting substantial domain shift when transferring from non-VPN to VPN traffic. \citeauthor{Kotak2025}~\cite{Kotak2025} model VPN-encrypted traffic as a time series of sliding-window packet statistics and apply InceptionTime for classification, treating the VPN tunnel as a time-series transformation problem. \citeauthor{Roy2022}~\cite{Roy2022} demonstrate that early classification from short packet prefixes, using only packet sizes and inter-arrival times, can reduce observation time while maintaining accuracy, a finding that motivates the SPLT representation used in this work.

\paragraph{Datasets and evaluation protocols}
Two widely used benchmarks for VPN traffic classification are ISCX-VPN-2016~\cite{DraperGil2016} and VNAT~\cite{Jorgensen2024}. Both collect non-VPN and VPN traffic in separate sessions. This design means that the non-VPN and VPN subsets may differ not only in encapsulation but also in session timing, user behavior, application versions, and network conditions. When a classifier trained on non-VPN data is tested on VPN data from a different session, any performance degradation conflates the encapsulation effect with these session-level differences. The dataset used in this work~\cite{Razooqi2026Pekar} takes a different approach: it provides matched pre- and post-VPN views of the same underlying flows captured simultaneously, eliminating session-level confounds and enabling a cross-domain evaluation that more directly isolates the effect of encapsulation on classification-relevant signal. The capture and matching design is described in \Cref{sec:dataset}.

\section{Dataset and Matched-Capture Design}
\label{sec:dataset}

The experiments use the dataset published by \citeauthor{Razooqi2026Pekar}~\cite{Razooqi2026Pekar}, a flow-level collection of WireGuard tunnel traffic captured over two sessions totaling approximately 80~hours of residential traffic from 10~end-user devices.

The distinguishing property of this dataset is its simultaneous paired-capture design. Pre-tunnel (inner) and post-tunnel (outer) PCAPs are recorded at the same time on the same residential edge: inner captures are taken on the WireGuard tunnel interface, while an inline network TAP mirrors the encrypted traffic to a dedicated capture host. Inner packets are matched to their WireGuard-encapsulated outer counterparts at the packet level, achieving a match ratio above 99.9\%. Application labels are assigned via deep packet inspection (NFStream~6.6.0~\cite{AOUINI2022108719} with nDPI~5.0~\cite{6906427}) on the inner side, where payloads are visible, and transferred to the outer view through this matching. The result is a dataset where every flow has both an unencapsulated view with DPI-assigned labels and an encrypted-side view derived from the matched tunnel packets, both describing the same flows at the same time.

The dataset comprises 226,454~flows (122,975 in session~1; 103,479 in session~2). After concatenating both sessions, each flow record is split into a non-VPN view (inner-side features) and a VPN view (encrypted-side features), both sharing the same application category label. Flows with no matched encrypted-side packets (33) are excluded, and categories with fewer than 200~samples are removed, reducing the count from 20 to 14 categories and from 226,454 to 226,281~flows per view. 
The distribution is heavily skewed: \textit{Web}~(48.9\%) and \textit{Network}~(37.3\%) together account for 86.2\% of samples, while the smallest categories have fewer than 300~samples. This imbalance motivates the use of balanced class weights during training and balanced accuracy / macro F1 as primary evaluation metrics.
Full collection, matching, and curation details are provided in~\cite{Razooqi2026Pekar}.


\section{Methodology}
\label{sec:methodology}

\subsection{Feature Representations}
\label{sec:features}

\paragraph{Flow-level statistical features (FlowFeatures)}
The baseline representation consists of 21~per-flow aggregates: bidirectional packet and byte counts, mean and standard deviation of packet size, bidirectional minimum/mean/standard-deviation/maximum inter-arrival times, flow duration, source-to-destination and destination-to-source byte counts, per-direction minimum/mean/standard-deviation/maximum inter-arrival times, and bidirectional packet and byte rates. Both views are computed from the same matched physical packets under a single statistical implementation, producing identically defined 21-dimensional vectors for the non-VPN and VPN domains. Undefined values (e.g., rates of zero-duration flows) are imputed with column medians fitted on the training view only. Because these statistics discard packet ordering, they provide a non-temporal baseline.

\paragraph{SPLT early-flow fingerprints}
The Sequence of Packet Length and Time (SPLT) fingerprint encodes the first $N{=}50$ matched packet pairs selected in inner-view order as a three-channel sequence of shape $50 \times 3$. The outer fingerprint contains observations of those same packets, ordered by their outer-view timestamps; this preserves physical packet correspondence without assuming that encapsulation preserves packet order. The three channels are: (1)~direction ($-1$~source-to-destination, $+1$~destination-to-source); (2)~packet size in bytes; and (3)~inter-arrival time in milliseconds, the latter two log-transformed as $\ln(1+x)$. Flows shorter than 50~packets are zero-padded to length~50, with the direction value~$0$ reserved for padding positions. Retaining packet order preserves directional handshake patterns, burst structures, and request--response rhythms that are characteristic of individual application categories. Crucially, packet sizes and inter-arrival times remain observable after VPN encapsulation, making SPLT a natural candidate for cross-domain transfer. These choices contrast order-discarding aggregates with early packet sequences, and fixed-position tree processing with local sequential processing of the same SPLT values.

\subsection{Classification Models}
\label{sec:models}

\paragraph{Classical baselines (RF and XGBoost)}
Random Forest and XGBoost serve as non-temporal baselines. Both are applied to FlowFeatures directly and to SPLT by flattening the $50 \times 3$ tensor into a 150-dimensional vector. RF uses 500 decision trees with balanced class weighting. XGBoost uses 500 estimators with maximum depth~6, subsampling ratio~0.8, and per-sample balanced weights. For SPLT input, flattening preserves packet order as fixed feature positions but removes the sequence-aware inductive bias that allows temporal models to exploit local patterns. This provides a controlled comparison for assessing whether temporal modeling helps explain any cross-domain advantage of CNN1D on SPLT.

\paragraph{Multi-scale CNN1D}
The model applies parallel 1D convolutional branches with different kernel sizes to detect application signatures at multiple temporal scales in the SPLT sequence. Three branches with kernels $k \in \{3, 7, 11\}$ and 64~filters each extract local burst patterns, medium-range request--response structures, and longer handshake sequences, respectively. Their outputs are concatenated and processed by multi-head attention (4~heads), a residual block, and a dense classification head. The model is trained with balanced class weights and sparse focal loss ($\alpha{=}0.25$, $\gamma{=}2.0$) using Adam. This configuration has 550,158 trainable parameters.

\subsection{Evaluation Protocol}
\label{sec:evaluation}

\paragraph{Cross-domain transfer setting}
As described in \Cref{sec:dataset}, each flow record is split into a non-VPN view (inner-side features with DPI labels) and a VPN view (encrypted-side features with the same labels transferred through matching). In the cross-domain setting, each model is trained on non-VPN flows and evaluated on the corresponding VPN view ($n{=}226{,}281$). This is a matched-view transfer benchmark: the training and test sets contain two representations of the same flows. The evaluation isolates how well each model generalizes across the representation shift introduced by encapsulation, rather than testing generalization to unseen flows. This same-flow design is deliberate; pair-disjoint evaluation, in which both views of a flow are held out together, is the complementary protocol for deployment-style unseen-flow generalization. No VPN samples are seen during training. For CNN1D, a stratified 10\% validation split is held out from the non-VPN data for early stopping; validation tensors use clean (unaugmented) sequences. With session-level confounds removed, differences among configurations can be interpreted more directly as sensitivity to encapsulation-induced representation shift. Uncertainty is quantified with 95\% percentile intervals from $B{=}1{,}000$ pair-level bootstrap resamples.

\paragraph{Metrics}
Four metrics are reported with bootstrap intervals: accuracy, balanced accuracy (mean per-class recall), macro F1 (unweighted mean of per-class F1), and weighted F1. Given the 86\% majority-class dominance, balanced accuracy and macro F1 are the primary comparison metrics. In addition, macro-averaged average precision (AP) over one-vs-rest precision--recall curves is reported as a point estimate to characterize threshold-independent classifier behavior under the class-imbalanced distribution.

\paragraph{Training details}
Two augmentation strategies are applied to SPLT training data only: random timestep masking ($p{=}0.05$) and additive Gaussian noise ($\mathcal{N}(0, 0.02)$). CNN1D uses early stopping and learning-rate reduction on plateau. All models use a fixed random seed. Classical models are implemented in scikit-learn and XGBoost; the CNN1D model uses TensorFlow/Keras. On the campaign host with 16~vCPUs and an RTX~6000 Ada GPU, recorded training times were 19/34~s for RF/XGBoost with FlowFeatures, 38/95~s for RF/XGBoost with SPLT, and 480~s for CNN1D. Online per-flow inference latency was not instrumented, so we make no real-time latency claim; the 50-packet prefix bounds the observation horizon, not compute latency.

\section{Results}
\label{sec:results}

\Cref{tab:cross_domain} reports cross-domain results. RF and XGBoost are evaluated on both feature representations; CNN1D is evaluated only on SPLT, since it is designed for sequential early-flow input rather than aggregate FlowFeatures. Under the matched-view transfer benchmark, models are trained exclusively on non-VPN flows and evaluated on the VPN view of the same underlying flows ($n{=}226{,}281$), with 95\% bootstrap confidence intervals for all threshold-dependent metrics.

\begin{table*}[!htp]
\centering
\caption{Matched-view cross-domain results: trained on the non-VPN view, evaluated on the VPN view of the same flows. Bold denotes the best value per column. 95\% bootstrap CIs ($B{=}1{,}000$) in brackets.}
\label{tab:cross_domain}
\begin{tabular}{@{}llcccc@{}}
\toprule
\textbf{Model} & \textbf{Features} & \textbf{Accuracy} & \textbf{Bal. Accuracy} & \textbf{Macro F1} & \textbf{Weighted F1} \\
\midrule
RF & FlowFeatures & 0.9680 {\scriptsize [0.9673, 0.9687]} & 0.8427 {\scriptsize [0.8376, 0.8477]} & 0.8262 {\scriptsize [0.8216, 0.8306]} & 0.9664 {\scriptsize [0.9657, 0.9672]} \\
XGBoost & FlowFeatures & 0.9421 {\scriptsize [0.9411, 0.9430]} & 0.9271 {\scriptsize [0.9230, 0.9312]} & 0.8635 {\scriptsize [0.8592, 0.8676]} & 0.9477 {\scriptsize [0.9468, 0.9485]} \\
RF & SPLT & 0.9396 {\scriptsize [0.9386, 0.9406]} & 0.6040 {\scriptsize [0.5974, 0.6101]} & 0.6364 {\scriptsize [0.6296, 0.6424]} & 0.9283 {\scriptsize [0.9271, 0.9296]} \\
XGBoost & SPLT & 0.9584 {\scriptsize [0.9576, 0.9592]} & 0.7534 {\scriptsize [0.7468, 0.7594]} & 0.7912 {\scriptsize [0.7851, 0.7968]} & 0.9559 {\scriptsize [0.9550, 0.9568]} \\
CNN1D & SPLT & \textbf{0.9682} {\scriptsize [0.9675, 0.9690]} & \textbf{0.9753} {\scriptsize [0.9733, 0.9773]} & \textbf{0.8932} {\scriptsize [0.8889, 0.8970]} & \textbf{0.9705} {\scriptsize [0.9699, 0.9712]} \\
\bottomrule
\end{tabular}

\end{table*}

Classical models on FlowFeatures retain useful cross-domain performance. XGBoost reaches a balanced accuracy of \XGBFlowFeaturesBalancedAccuracy{} and RF \RFFlowFeaturesBalancedAccuracy{} on the same 21-dimensional feature space; as \Cref{sec:discussion} shows, most of the RF deficit is concentrated in a single category rather than spread across classes.

Classical models yield lower balanced accuracy with flattened SPLT: RF reaches \RFSPLTBalancedAccuracy{} and XGBoost \XGBSPLTBalancedAccuracy.

CNN1D on SPLT achieves the best value in every reported metric: accuracy \CNNOneDSPLTAccuracy, balanced accuracy \CNNOneDSPLTBalancedAccuracy, macro F1 \CNNOneDSPLTMacroFOne, and weighted F1 \CNNOneDSPLTWeightedFOne, with narrow bootstrap confidence intervals (macro F1 width 0.008). This is achieved without any VPN data during training. Its accuracy is similar to RF with FlowFeatures (\RFFlowFeaturesAccuracy), with overlapping confidence intervals, while its balanced accuracy exceeds the best classical configuration by 0.048.

 \begin{figure}[t]
    \centering
    \includegraphics[width=.9\columnwidth]{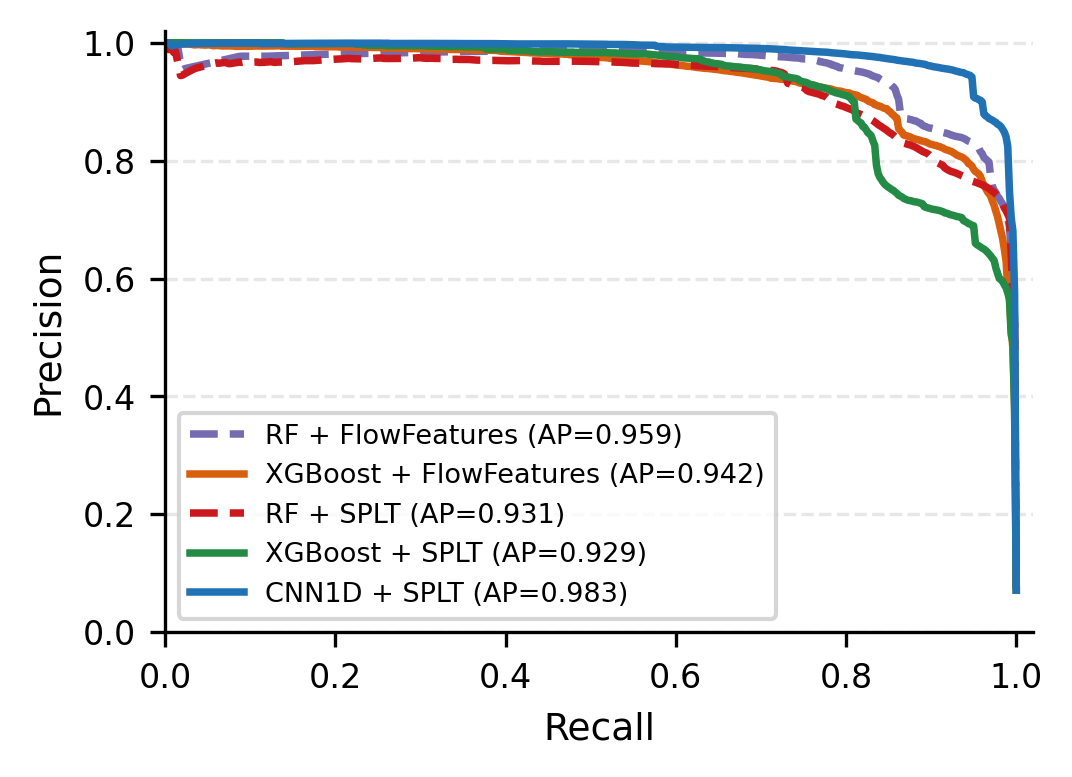}
    \vspace{-.5cm}
    \caption{Macro-averaged precision--recall curves for five
             cross-domain configurations. CNN1D\,+\,SPLT (AP\,=\,0.983)
             sustains precision above 0.97 up to recall\,$\approx$\,0.86.
             The classical configurations occupy a band between
             AP\,=\,0.929 and 0.959, with XGBoost\,+\,SPLT dropping
             earliest, above recall\,$\approx$\,0.8.}
    \label{fig:pr_macro}
  \end{figure}

\Cref{fig:pr_macro} shows the macro-averaged precision--recall curves for all five cross-domain configurations. CNN1D with SPLT achieves the highest average precision (AP\,=\,0.983), sustaining precision above 0.97 up to recall\,$\approx$\,0.86 and remaining the leading curve through most of the recall range. Among the classical configurations, RF attains a higher AP than XGBoost on both representations (FlowFeatures: 0.959 vs.\ 0.942; SPLT: 0.931 vs.\ 0.929), even where its single-threshold metrics are lower, a probability-ranking property examined in \Cref{sec:discussion_interaction}. Both SPLT tree configurations show an earlier precision drop than their FlowFeatures counterparts.

Taken together, the cross-domain results reveal a clear interaction between feature representation and model family: neither alone determines transfer success. This interaction is analyzed in detail in \Cref{sec:discussion}.

\section{Discussion}
\label{sec:discussion}

\subsection{Per-Class Cross-Domain Transfer}
\label{sec:perclass}

\begin{figure*}[t]
  \centering
  \includegraphics[width=.75\textwidth]{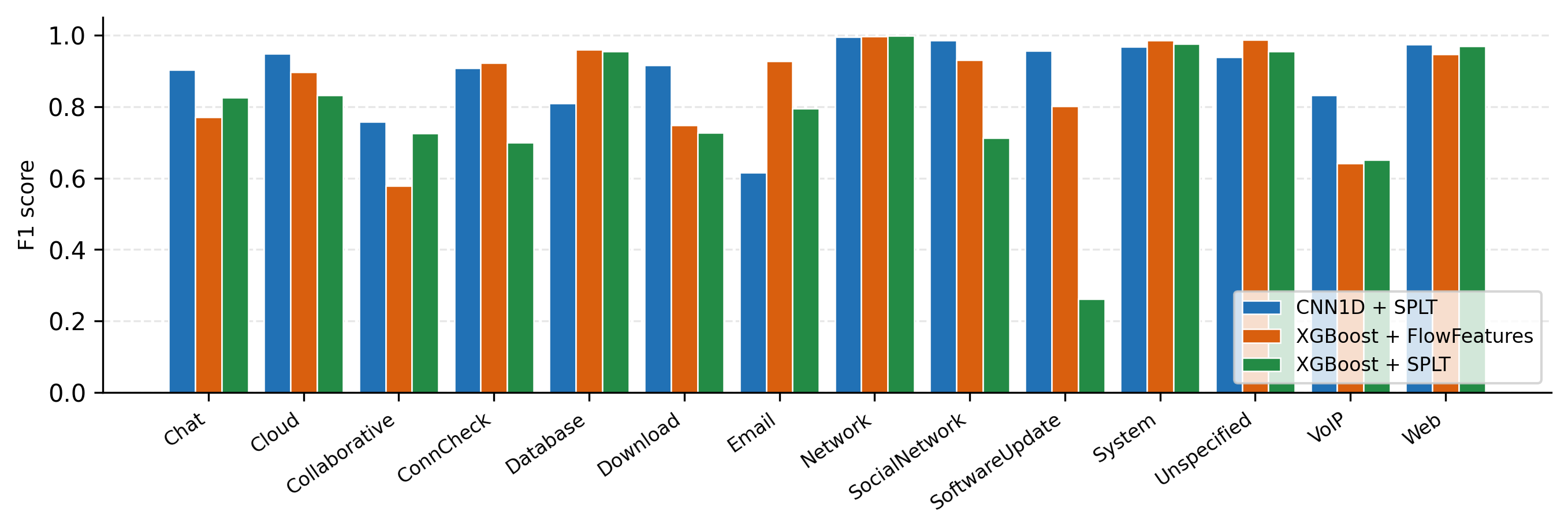}
  \vspace{-.5cm}
  \caption{Per-class F1 scores for the three key cross-domain configurations.
           CNN1D\,+\,SPLT (blue) leads on seven of the twelve minority
           categories; XGB\,+\,SPLT (green) falls to F1\,=\,0.26 on SoftwareUpdate
           while CNN1D dips only on Email (F1\,=\,0.61),
           illustrating the representation--model interaction.}
  \label{fig:perclass_f1}
\end{figure*}

Cross-domain transfer is not uniform across application categories, and the per-class breakdown in \Cref{fig:perclass_f1} exposes where each configuration succeeds or fails. Network and Web, together accounting for 86.2\% of the test set, achieve F1 above 0.94 under all three configurations, suggesting that their dominant volume and distinctive aggregate profiles provide sufficient signal regardless of how the feature is encoded or which model is used.

The twelve minority categories show a pronounced representation--model interaction. CNN1D with SPLT leads on seven of them, with the largest margins over XGBoost with FlowFeatures on temporally structured classes: VoIP (+0.19~F1), Collaborative (+0.18), Download (+0.17), SoftwareUpdate (+0.16), and Chat (+0.13). XGBoost with FlowFeatures remains stronger where whole-flow statistics are discriminative enough, such as Database (F1\,=\,0.96), Unspecified (0.99), and Email (0.93). The starkest flattening failure is SoftwareUpdate: XGBoost with SPLT recalls only 15\% of its flows (F1\,=\,0.26), while CNN1D reaches an F1 of 0.96 on the identical SPLT input. Conversely, CNN1D's single weak category is Email (F1\,=\,0.61): it recalls 99\% of Email flows but at a precision of 0.45, over-predicting the category.

For CNN1D with SPLT, errors concentrate on a few category pairs. The largest confusion is Web being misclassified as Collaborative (3,515 flows; 3.2\% of Web), with Collaborative misclassified as Web in return (4.0\%). This reciprocal confusion may reflect similar short burst patterns, but confirming the mechanism would require a feature-level ablation. The Email precision loss traces almost entirely to Unspecified flows misclassified as Email (8.0\% of Unspecified).

\subsection{Representation--Model Interaction}
\label{sec:discussion_interaction}

The cross-domain results in \Cref{tab:cross_domain} reveal that no feature representation dominates across model families: FlowFeatures transfer well under both tree ensembles, the identical SPLT input spans the worst and best configurations depending on the model, and the ranking of RF versus XGBoost depends on the metric. This interaction warrants closer examination.

Balanced accuracy exceeds macro F1 for every configuration. This pattern is consistent with balanced class weighting and focal loss favoring minority recall while precision drops on categories whose profiles overlap larger classes (e.g., XGBoost with FlowFeatures: VoIP precision 0.51 at recall 0.86; Collaborative 0.45 at 0.79). CNN1D attains both higher precision and higher recall than XGBoost with FlowFeatures on most minority categories, suggesting that sequence-aware processing may retain discriminative temporal information absent from the aggregate representation.

RF with FlowFeatures trails XGBoost in balanced accuracy (0.8427 vs.\ 0.9271) yet exceeds it in accuracy (0.9680 vs.\ 0.9421) and average precision (0.959 vs.\ 0.942). The per-class breakdown resolves this apparent contradiction: RF's deficit is almost entirely one category. On System, RF recalls 1.4\% of flows at a precision of 1.00 (F1\,=\,0.03), while XGBoost reaches an F1 of 0.99; excluding System, the two models' mean per-class recall differs by only 0.016. The same System failure appears when RF is applied to SPLT (F1\,=\,0.02), suggesting a model-specific decision behavior rather than a representation-specific effect. RF's high average precision further shows that its class-probability ordering remains useful even though its argmax predictions almost never select System, a distinction that single-threshold metrics cannot expose.

The most striking result is the gap among models using SPLT: RF and XGBoost reach balanced accuracy of 0.6040 and 0.7534, against 0.9753 for CNN1D. This indicates a representation--model mismatch rather than an absence of signal: SoftwareUpdate and SocialNetwork reach F1 of 0.25/0.26 and 0.33/0.71 under the flattened tree models, but 0.96 and 0.99 under CNN1D. A plausible explanation is that flattening encodes each packet position as an independent feature, so the position-specific thresholds learned by tree models may transfer poorly when encapsulation shifts packet sizes and timing. CNN1D may be more robust because its convolutional filters operate on local neighborhoods rather than absolute positions; an ablation is needed to verify this mechanism.

Because the inner and outer views describe the same flows, differences among the cross-domain results can be interpreted as sensitivity to the encapsulation-induced representation shift without conflating separate-session variation. The CNN1D balanced accuracy of 0.98 is best read as robustness to encapsulation-induced distortion of early-flow temporal patterns, not deployment-style generalization to unseen flows.

\section{Conclusion}
\label{sec:conclusion}

This paper presented a matched-view cross-domain transfer benchmark for application category classification on WireGuard VPN traffic, built on a matched-capture dataset where pre- and post-tunnel views of the same flows are recorded simultaneously, eliminating the session-level confounds in prior datasets that collect VPN and non-VPN traffic separately. Under this benchmark, transfer success depends jointly on the feature representation and the model: tree ensembles achieve higher balanced accuracy and macro F1 with FlowFeatures than with flattened SPLT, while CNN1D achieves the strongest result with SPLT (balanced accuracy 0.98) without any VPN data during training. This interaction implies that VPN classification systems should match the representation to the model's inductive bias rather than optimizing either in isolation.

The evaluation covers a single VPN protocol (WireGuard), a single residential capture site, and two sessions from 10~devices, with a heavily skewed class distribution; the matched-view benchmark also isolates transfer across the encapsulation-induced representation shift rather than deployment-style generalization to unseen flows. These constraints bound the generalizability of the specific numerical results, though the matched-capture methodology itself is protocol-independent. Ablation of SPLT design choices, additional VPN protocols, unseen-flow and cross-session evaluation, and adapted literature baselines are directions for future work.

\section*{Reproducibility}
All preprocessing, training, evaluation, and plotting code is available~\cite{FlowFrontiersCode}, together with evaluation artifacts; the dataset is published separately~\cite{Razooqi2026Pekar}.

\section*{Acknowledgement}

Supported by the CELTIC-NEXT project Robust and AI Native 6G for Green Networks (RAI6-Green, C2023/1-9), funded by the National Research, Development and Innovation Fund of Hungary under Grant-2024-1.2.6-EUREKA-2024-00009.

\printbibliography

\end{document}